\documentclass{aa}  

\usepackage{graphicx}
\usepackage{txfonts}
\usepackage{enumitem, array}
\usepackage{subcaption}

\newcommand{\nscool}{\texttt{NSC\textsc{ool}}\,}
\newcommand{\kteff}{$kT^{\infty}_{\mathrm{eff}}$\,}
\newcommand{\nh}{$N_\mathrm{H}$\,}
\newcommand{\Qimp}{$Q_\mathrm{imp}$}
\newcommand{\Qsh}{$Q_\mathrm{sh}$}
\newcommand{\rhosh}{$\rho_\mathrm{sh}$}
\newcommand{\mdot}{$\langle \dot{\mathrm{M}} \rangle$\,}
\newcommand{\mxb}{MXB 1659$-$29 }
\newcommand{\fbol}{$F_\mathrm{bol}$}

\begin{document}

\title{Consistent accretion-induced heating of the neutron-star crust in \mxb during two different outbursts}

\author{A. S. Parikh\inst{1}
 \and
 R. Wijnands\inst{1}
 \and
 L. S. Ootes\inst{1}
 \and
 D. Page\inst{2}
 \and
 N. Degenaar\inst{1}
 \and
 A. Bahramian\inst{3}
 \and
 E. F. Brown\inst{4}
 \and
 E. M. Cackett\inst{5}
 \and
A. Cumming\inst{6}
 \and
 C. Heinke\inst{7}
 \and
 J. Homan\inst{8,9}
 \and
 A. Rouco Escorial\inst{1}
 \and
  M. J. P. Wijngaarden\inst{10}
 }

\institute{Anton Pannekoek Institute for Astronomy, University of Amsterdam, Postbus 94249, 1090 GE Amsterdam, The Netherlands\\
\email{a.s.parikh@uva.nl}
\and
Instituto de Astronom\'{i}a, Universidad Nacional Aut\'{o}noma de M\'{e}xico, Mexico D.F. 04510, Mexico
\and
International Centre for Radio Astronomy Research – Curtin University, GPO Box U1987, Perth, WA 6845, Australia
\and
Department of Physics and Astronomy, Michigan State University, 567 Wilson Road, East Lansing, MI 48864, USA
\and
Department of Physics \& Astronomy, Wayne State University, 666 W. Hancock Street, Detroit, MI 48201, USA
\and
Department of Physics and McGill Space Institute, McGill University, 3600 Rue University, Montreal, QC H3A 2T8, Canada
\and
Department of Physics, University of Alberta, CCIS 4-181, Edmonton, AB T6G 2E1, Canada
\and
SRON, Netherlands Institute for Space Research, Sorbonnelaan 2, 3584 CA Utrecht, The Netherlands
\and
Eureka Scientific, Inc., 2452 Delmer Street, Oakland, CA 94602, USA
\and           
Mathematical Sciences, University of Southampton, SO17 1BJ, Southampton, United Kingdom             }

\titlerunning{NS crust cooling in \mxb}
\authorrunning{A. S. Parikh et al.}

   \date{Received x; accepted x}

 
  \abstract
{
Monitoring the cooling of neutron-star crusts heated during accretion outbursts allows us to infer the physics of the dense matter present in the crust. We examine the crust cooling evolution of the low-mass X-ray binary \mxb up to $\sim$505 days after the end of its 2015 outburst (hereafter outburst II) and compare it with what we observed after its previous 1999 outburst (hereafter outburst I) using data obtained from the {\it Swift}, {\it XMM-Newton}, and {\it Chandra} observatories. The observed effective surface temperature of the neutron star in \mxb dropped from $\sim$92 eV to $\sim$56 eV from $\sim$12 days to $\sim$505 days after the end of outburst II. The most recently performed observation after outburst II suggests that the crust is close to returning to thermal equilibrium with the core. We model the crust heating and cooling for both its outbursts collectively to understand the effect of parameters that may change for every outburst (e.g., the average accretion rate, the length of outburst, the envelope composition of the neutron star at the end of the outburst) and those which can be assumed to remain the same during these two outbursts (e.g., the neutron star mass, its radius). Our modelling indicates that all parameters were consistent between the two outbursts with no need for any significant changes. In particular, the strength and the depth of the shallow heating mechanism at work (in the crust) were inferred to be the same during both outbursts, contrary to what has been found when modelling the cooling curves after multiple outburst of another source, MAXI J0556$-$332. This difference in source behaviour is not understood. We discuss our results in the context of our current understanding of cooling of accretion-heated neutron-star crusts, and in particular with respect to the unexplained shallow heating mechanism.

}


   \maketitle

%

\section{Introduction}
\label{sec_introduction}
The density in the crust of a neutron star (NS) increases by eight orders of magnitude over $\sim$1 km, from the outer crust of the NS ($\rho\sim10^{6}$ g cm$^{-3}$) to the crust-core boundary ($\rho \sim 1.5 \times 10^{14}$ g cm$^{-3}$). Thus, NS crusts provide an excellent opportunity to study the behaviour of dense matter over a large density range. One of the ways in which we can do this is by studying the cooling of accretion-heated NS crusts. Several NSs in low-mass X-ray binaries (LMXBs; binary systems wherein the donor is typically a sub-Solar star) experience transient outbursts during which matter from a disk around the NS is accreted onto its surface. This results in compression-induced exothermic nuclear reactions in the crust that can disrupt the crust-core thermal equilibrium \citep[][]{haensel1990non,haensel2003nuclear,haensel2008models,steiner2012deep}. In these transient systems, the outbursts are separated by periods of quiescence during which no (or only very little) matter accretes onto the NS surface and as a result no (significant) heating by compression-induced reactions occur. When such accretion outbursts have halted, the crust begins to cool in order to reinstate thermal equilibrium with the core (if this equilibrium was disrupted during the outburst). Monitoring this cooling has given us an invaluable insight into the properties of matter over the high densities that occur in the NS crust, although many uncertainties remain (e.g., \citeauthor{shternin2007neutron} \citeyear{shternin2007neutron}; \citeauthor{brown2009mapping} \citeyear{brown2009mapping}; see \citeauthor{meisel2018nuclear} \citeyear{meisel2018nuclear}, for a review of the theoretical advances).

Currently, crust cooling curves have been obtained for nine NSs in LMXBs \citep[see][for an observational review]{wijnands2017review}. Modelling these observed crust cooling curves with theoretical models indicated that, besides the deep crustal heating mechanism (occurring deep in the crust, at densities of $\rho \sim10^{12}$--$10^{13}$ g cm$^{-3}$), an additional, unknown crustal heat source should be active during the accretion outbursts of most sources to explain their early cooling evolution \citep[e.g.,][]{brown2009mapping,degenaar2014probing,parikh2017potential,wijnands2017review}. This heat source is typically referred to as the shallow heating mechanism because, as the name suggests, it occurs at a shallower depth in the NS crust (i.e., at lower densities: $\rho \sim10^{8}$--$10^{10}$ g cm$^{-3}$) than the deep crustal heating. 

MXB 1659$-$29 was discovered in 1976 \citep{{lewin1976xray}} as a transient LMXB that exhibited type-I X-ray bursts (which are caused by a runaway thermonuclear burning process on the NS surface) which established the NS nature of the accretor. The source also was found to exhibit eclipses, lasting $\sim$900 s, during its $\sim$7.1 hours orbit \citep{cominsky1984discovery,jain2017indication,iaria2018a}. This outburst lasted $\sim$2--2.5 years \citep{wijnands2003chandra}. There were no follow-up observations to study the crust cooling of \mxb after this outburst. A second accretion outburst from the source was detected in 1999  \citep[][]{zand1999v2134} that lasted $\sim$2.5 years as well \citep{wijnands2002burst}. Post-outburst observations (i.e., when the accretion had halted) using {\it Chandra} and {\it XMM-Newton} found a cooling NS crust \citep[][this outburst is further referred to as outburst I as it was the first outburst in \mxb after which crust cooling was investigated]{wijnands2003chandra,wijnands2004monitoring,cackett2006cooling,cackett2008cooling,cackett2013change}. It was the second source, after KS 1731-260 \citep{wijnands2003chandra}, for which such crust cooling was established. \mxb was further monitored, up to $\sim$11 years after the end of this outburst \citep[e.g.,][]{cackett2010continued,ootes2016,merritt2016thermal} until the source displayed a new outburst in 2015 \citep{negoro2015mxb}. The crust cooling of the NS in \mxb monitored over this period, in addition to a similar monitoring of KS 1731$-$260, led to a significant leap in our understanding of the physics of the NS crust. Contrary to original expectation \citep{schatz2001endpoint}, the NS crust in both sources was found to have a high thermal conductivity because it is likely a highly structured crystallised crust \citep[with a low number of impurities;][]{shternin2007neutron,brown2009mapping}. 

Following the long-term cooling of \mxb allowed for the NS dense matter behaviour to be probed at all depths, from the topmost layers of the crust \citep{turlione2015quiescent,horowitz2015disordered,deibel2017late} to the core \citep{cumming2017lower,brown2018rapid}, assuming we observed crust-core equilibrium again at the end of the available cooling curve \citep{cackett2013change}. However, this assumption may not be entirely valid since the interpretation of the results obtained during the last observations ({\it Chandra} observation IDs [obsIDs]: 13711 and 14453, carried out $\sim$3 days apart) after this outburst is ambiguous \citep[see][for details]{cackett2013change}: either the decrease in count rate between these observations and the previous one could be due to a further cooling of the crust, or due to an increased internal absorption (e.g., due to an increase in the height of the outer accretion disk) and the crust had not further cooled. Neither scenario could be confirmed using additional observations since soon after the last observation the source exhibited its next accretion outburst. We assume that the most likely scenario is that of the increase in internal absorption meaning that the crust did not cool further. Therefore, we do not use these last two {\it Chandra} observations after outburst I in our study and we assume that the constant (plateau) level observed near the end of the cooling curve after outburst I is representative of the crust returning to thermal equilibrium with the core. Future quiescent observations after the end of its most recent outburst may help break the ambiguity of the interpretation of this last set of observations after outburst I (see Section \ref{sec_disc}).

MXB 1659$-$29 exhibited a new accretion outburst in 2015 \citep[further referred to as outburst II;][]{negoro2015mxb} which lasted $\sim$1.7 years and the source transitioned back to quiescence in 2017 March \citep{parikh2017neutron}. After the end of this outburst, we started a sequence of {\it XMM-Newton} and {\it Chandra} observations to obtain a second crust cooling curve of this source. The early, preliminary cooling results, up to $\sim$26 days after the end of the outburst (and thereby probing only the upper layers of the crust), were reported by us in \citet{wijngaarden2017window}. Here, we report on observations up to $\sim$505 days after the end of this outburst which allowed us to probe the physics of the deeper crust.


\section{Observations, Data Analysis, and Results}

\mxb is viewed at a high inclination \citep[$i \sim 69$--77$^{\circ}$;][]{iaria2018a,ponti2018measuring} and eclipses are observed. To obtain the true intrinsic luminosity of the source (both during the outburst to estimate \mdot and during quiescence to determine the effective NS surface temperature [\kteff]) only the non-eclipsing `persistent' data should be used. Therefore,  all data were corrected for  eclipses using the ephemeris reported by \citet{iaria2018a} by artificially reducing the exposure time corresponding to the number of the eclipses that occur during an observation. The eclipse lasts for $\sim$900 seconds of the $\sim$7.1 hour orbital period of MXB 1659$-$29.

\begin{table*}
\centering
\caption{The log of the quiescent observations used in our paper and the results of the spectral fitting are shown.$^\text{a}$}
\label{tab_log}
\small{

\begin{tabular}{p{0.1cm}p{1.8cm}>{\centering\arraybackslash}p{1.3cm}>{\centering\arraybackslash}p{1.7cm}>{\centering\arraybackslash}p{0.65cm}>{\centering\arraybackslash}p{2.1cm}>{\centering\arraybackslash}p{3.45cm}>{\centering\arraybackslash}p{1cm}>{\centering\arraybackslash}p{1.4cm}>{\centering\arraybackslash}p{1cm}}
\hline
& Observatory & ObsID & Date & MJD &Exposure & Count		& $kT_\text{eff}^\infty$& $F_\text{X}^\text{d}$ & $L_\text{X}^\text{d}$ \tabularnewline
& 					&&			&&time$^\text{b}$& Rate$^\text{c}$ & (eV)				& ($\times 10^{-14}$			& ($\times 10^{32}$ \tabularnewline
& 					&&		& &(ksec)	& ($\times 10^{-3}$ counts s$^{-1}$)& & erg cm$^{-2}$ s$^{-1}$) & erg s$^{-1}$)\tabularnewline
\hline
\tabularnewline
\multicolumn{10}{>{\centering\arraybackslash}p{16cm}}{{\it After Outburst I}} \tabularnewline
1 & {\it Chandra} 	& 2688 		& 2001 Oct 15 & 52197.7	& 17.9 			& 52.1$\pm$1.7					& 111.1$\pm$1.3  & 35.4$\pm$1.9 &   34.3$\pm$1.8	\tabularnewline
2 & {\it Chandra} 	& 3794 		& 2002 Oct 26 & 52563.0 	& 26.2 			& 9.6$\pm$0.6					& 79.5$\pm$1.6   &  7.5$\pm$0.8 &    7.3$\pm$0.8	\tabularnewline
3 & {\it XMM-Newton} 	& 0153190101 	& 2003 Mar 13 & 52711.6 	& 8.2, 31.1, 13.1 	& 11.3$\pm$1.4, 2.6$\pm$0.3, 2.4$\pm$0.5	& 73.0$\pm$1.9   &  5.0$\pm$0.7 &    4.8$\pm$0.6	\tabularnewline
4 & {\it Chandra} 	& 3795 		& 2003 May 9 & 52768.7	& 25.6 			& 3.7$\pm$0.4					& 67.8$\pm$2.1   &  3.4$\pm$0.6 &    3.3$\pm$0.6	\tabularnewline
5 & {\it Chandra} 	& 5469/6337$^\text{e}$ 	& 2005 Jul 15 & 53566.4 	& 45.5 			& 1.0$\pm$0.2 				& 55.5$\pm$2.4   &  1.2$\pm$0.3 &    1.2$\pm$0.3	\tabularnewline
6 & {\it Chandra} 	& 8984 		& 2008 Apr 27 & 54583.8 	& 26.6 			& 0.9$\pm$0.2					& 54.8$\pm$3.2   &  1.1$\pm$0.4 &    1.1$\pm$0.4	\tabularnewline
\tabularnewline
\multicolumn{10}{>{\centering\arraybackslash}p{16cm}}{{\it After Outburst II}} \tabularnewline   
1 & {\it Swift} 	& Interval 1$^\text{f}$ & 2017 Mar 10 	& 57822.0	& 6.6	 		& 1.8$\pm$0.5			&  91.5$\pm$8.8   & 13.4$\pm$6.3 &   13.0$\pm$6.1 	\tabularnewline
2 & {\it XMM-Newton} 	& 0803640301 	& 2017 Mar 23 & 57835.8	& 5.3, 22.4, 20.5 	& 32.4$\pm$2.7, 6.6$\pm$0.6, 6.2$\pm$0.6	&  87.9$\pm$1.4   & 11.4$\pm$1.0 &   11.1$\pm$0.9 	\tabularnewline
3 & {\it Chandra} 	& 19599 	& 2017 Apr 25 & 57868.0	& 26.4 			& 5.7$\pm$0.5					&  82.7$\pm$2.0   &  8.8$\pm$1.2 &    8.6$\pm$1.2 	\tabularnewline
4 & {\it Chandra} 	& 19600 	& 2017 Jul 3 & 57937.6	& 25.2 			& 3.1$\pm$0.4					&  74.8$\pm$2.5   &  5.5$\pm$1.0 &    5.3$\pm$1.0 	\tabularnewline
5 & {\it XMM-Newton} 	& 0803640401 	& 2017 Aug 22 & 57987.3 	& 2.2, 14.3, 21.2 	& 20.0$\pm$0.3, 3.1$\pm$0.5, 1.7$\pm$0.3	&  75.1$\pm$2.4   &  5.7$\pm$0.9 &    5.6$\pm$0.9 	\tabularnewline
6 & {\it Chandra} 	& 19601 	& 2018 Feb 2 & 58151.5 	& 26.4 			& 1.4$\pm$0.2					&  66.0$\pm$3.0   &  3.0$\pm$0.8 &    2.9$\pm$0.8 	\tabularnewline
7 & {\it Chandra} 	& 19602 	& 2018 Jul 15 & 58314.4	& 38.2			& 0.5$\pm$0.1						&  56.3$\pm$4.2   &  1.3$\pm$0.6 &    1.3$\pm$0.6 	\tabularnewline\hline

\hline
\multicolumn{10}{p{18.7cm}}{$^\text{a}$All errors are stated for the 90 per cent confidence level. The \nh is fixed to 3.4$\times 10^{21}$ cm$^{-2}$. For the {\it XMM-Newton} spectra, all the model parameters (except the normalisation constant) were tied between the three detectors for a given observation. The {\it XMM-Newton} exposure times and count rates are displayed as `pn, MOS1, MOS2'. We have assumed that the source is at a distance of 9 kpc.} \tabularnewline
\multicolumn{10}{p{17.7cm}}{$^\text{b}$The exposure times listed are the effective ones, after the data were modified for background flaring and eclipses.} \tabularnewline
\multicolumn{10}{p{17.7cm}}{$^\text{c}$The effective count rates (0.3--10 keV) have been modified for background flaring and eclipses.} \tabularnewline
\multicolumn{10}{p{16cm}}{$^\text{d}$The flux and luminosity correspond to the unabsorbed flux and luminosity for the 0.5--10 keV energy range.} \tabularnewline
\multicolumn{10}{p{16cm}}{$^\text{e}$The spectra from these two {\it Chandra} observations have been combined, see Sect. \ref{sec_chandra}.}\tabularnewline
\multicolumn{10}{p{16cm}}{$^\text{f}$Five {\it Swift}/XRT observations have been combined as Interval 1, see Sect. \ref{sec_swift}.}\tabularnewline
\end{tabular}
}
\end{table*}

\subsection{Light curves}
\label{sect_lc}
Outburst I was observed using the All-Sky Monitor (ASM) aboard the {\it Rossi X-ray Timing Explorer} ({\it RXTE}). Data from the more sensitive {\it RXTE}/Proportional Counting Array (PCA) was used to track the end of outburst I. Outburst II was observed using the Gas Slit Camera (GSC) on board the {\it Monitor of All-Sky X-ray Image} ({\it MAXI}) and the X-ray Telescope (XRT) on board the {\it Neil Gehrels Swift Observatory}. 

The 2--10 keV light curves\footnote{The one-day binned light curves were obtained from the respective archives:\\ASM: http://xte.mit.edu/asmlc/One-Day.html\\ {\it MAXI}: http://maxi.riken.jp/mxondem/} obtained from the ASM and {\it MAXI} instruments were rebinned, with a maximum of 4 days per bin, to increase the data statistics. The data were further filtered such that points having error bars >0.27 counts s$^{-1}$ and  >0.28 counts s$^{-1}$ were removed from the ASM and {\it MAXI} data, respectively. Examining the ASM light curve indicated that even when outburst I was over (based on the more sensitive PCA data as reported in a later paragraph and the first quiescent {\it Chandra} observation as discussed in Section \ref{sec_chandra}) the source was detected at $\lesssim$0.7 counts s$^{-1}$. Therefore, these ASM detections are not real and we removed all ASM data <0.7 counts s$^{-1}$  for these points. All {\it MAXI} data <0.005 counts s$^{-1}$  were removed to ensure that we only consider observations during which the source was conclusively detected. 

Our raw {\it Swift}/XRT data (obsID: 00034002001--00034002087)\footnote{https://heasarc.gsfc.nasa.gov/cgi-bin/W3Browse/swift.pl} were processed using \texttt{xrtpipeline} (\texttt{HEASOFT}; v6.22). The background corrected light curve was generated using \texttt{XSelect} (v2.4). A circular source region having a radius of 50\arcsec\, centred on the source position was used \citep{wijnands2003chandra}. As background region we used an annulus (again centred on the source position) with an inner and outer radius of 175\arcsec\, and 300\arcsec, respectively. The data were corrected for pile-up when necessary. All type-I X-ray bursts (found by visually inspecting the light curves) were removed from the data. In addition to eclipses, \mxb also exhibits strong dipping behaviour preceding the eclipses \citep{cominsky1984discovery}. Only when the data quality was high (e.g., during the outburst observations using the XRT), these dips could be discerned from the persistent flux in the light curves.  The data were corrected for this dipping (by discarding the intervals) whenever they were clearly visible (by eye) in the light curves. Additionally, \citet{wijnands2003chandra} also observed dipping when the source was in quiescence. Thus, this dipping behaviour contributes a systematic source of uncertainty in quiescent observations (which is difficult to model) .

Determining the date of the end of the outburst is important in our cooling model (Section \ref{sect_nscool}). The source was last detected by the XRT during outburst II on MJD 57799.8. During the subsequent observation $\sim$20 days later the source was detected in quiescence at a factor $\sim$600 lower count rate. We assume that the outburst ended on MJD 57809.7, as determined by linearly interpolating between the date on which the source was last detected in outburst and subsequently detected at a factor $\sim$600 lower level (in quiescence) for the first time.\footnote{We have also modelled the observed cooling evolution (see later sections) assuming the end of outburst was the last day on which the source was detected in outburst (MJD 57799.8) or, alternatively, the first day it was detected at a factor $\sim$600 lower level (MJD 57819.6), in quiescence. These two options constitute the two most extreme (albeit unlikely) possibilities for the exact end date of the outburst. We find that changing the end of outburst date does not change the broad physical interpretation of our results.}

We also determine the end of outburst I in the same manner, for consistency. Since the ASM is not very sensitive at the low count rates near the end of an outburst, we have used the data obtained from the PCA on board {\it RXTE} near the end of the outburst as reported by \citet{wijnands2002burst}. They found that the source was detected by the PCA on MJD 52159 at $\sim$5 mCrabs. Assuming, 1 Crab (2--60 keV) =  2.4$\times$10$^{-8}$ erg cm$^{-2}$ s$^{-1}$ (0.5--10 keV) we find that \mxb was detected by the PCA at $\sim$1.2$\times$10$^{-10}$ erg cm$^{-2}$ s$^{-1}$. The source was not detected on MJD 52166 (with the upper limit corresponding to $\lesssim$1 mCrab [2--60 keV] i.e., $\lesssim$2.4$\times$10$^{-11}$ erg cm$^{-2}$ s$^{-1}$ [0.5--10 keV]). Therefore, linearly interpolating between these dates we calculate MJD 52162 to represent the end of outburst I. This is different from the end of outburst date assumed by \cite{cackett2008cooling} as they assumed that the last day on which the source was detected in outburst corresponded to the end of the outburst (MJD 52159.5). We have updated this assumption here, to be consistent with our analysis of outburst II. 

The light curves described here are presented as bolometric flux curves in Figure \ref{fig_lc} (see Section \ref{sect_nscool}, for details). Outburst I lasted for $\sim$2.5 years whereas outburst II lasted for $\sim$1.7 years. Outburst I transitioned smoothly from a constant flux level during the outburst to a rapid decay near the end of the outburst. Outburst II exhibited a lot more variability during the last $\sim$5 months and did not transition to the outburst decay as smoothly as observed for outburst I.

\subsection{Post-outburst spectral analysis}
\label{sect_spectral_analysis} 
We present five new observations of \mxb after the end of outburst II, in addition to the two intervals reported by \citet{wijngaarden2017window}. So far, \mxb has been observed twice using {\it XMM-Newton} and four times using {\it Chandra}, up to $\sim$505 days after the end of this outburst. The early post-outburst II cooling evolution could be constrained by combining several {\it Swift}/XRT observations. In Section \ref{sect_nscool},  we modelled the results obtained from the quiescent observations after both outbursts collectively to obtain the best constraints on the crustal physics. Therefore, for uniformity, we have also reanalysed all the observations after the end of outburst I \citep{wijnands2003chandra,wijnands2004monitoring,cackett2006cooling,cackett2008cooling,cackett2013change},  in the same way as the observations performed after outburst II. The log of all the observations used in our paper is shown in Table \ref{tab_log}. 

\subsubsection{\textit{Swift}/XRT}
\label{sec_swift}
We combined five observations (obsID: 00034002072--00034002076, as Interval 1) taken $\sim$10--18 days after the end of outburst II. The count rates from these observations were consistent with one another. The Photon Counting mode (two-dimensional imaging) event files from these observations were combined and the count rate and spectrum were extracted using a circular source region having a radius of 20\arcsec centred on the source position. The same background region as used for the light curve extraction was used (Section \ref{sect_lc}). The ancillary response file was constructed using the \texttt{xrtmkarf} tool. The response matrix file swxpc0to12s6${\_}$20130101v014.rmf, as indicated by the \texttt{xrtmkarf} tool, was used.

\subsubsection{\textit{XMM-Newton}}
\mxb was observed once after the end of outburst I and twice after the end of outburst II using all three {\it XMM-Newton} European Photo Imaging Cameras (EPIC) -- pn, MOS1, and MOS2. The source was too weak to be detected significantly by the Reflection Grating Spectrometer, therefore, data from this instrument are not used. We do not use data from the Optical Monitor instrument as these are not useful for our cooling studies. The raw data\footnote{Obtained using the {\it XMM-Newton} archive: http://nxsa.esac.esa.int/nxsa-web/} were processed using the Science Analysis System (\texttt{SAS}; v16.1). The data (in the 10--12 keV energy range for the pn detector and >10 keV for the MOS detectors) were examined for background flares. To discard these flares, data exceeding >0.25--0.3 counts s$^{-1}$ and >0.2--0.3 counts s$^{-1}$ were removed from the appropriate pn and MOS observations, respectively. 

The source region used for the light curve and spectral extraction was calculated using the \texttt{eregionanalyse} tool to optimise the signal-to-noise ratio. Circular source regions (centred on the source position) having radii of 14.5\arcsec--18\arcsec\, and 12\arcsec--18\arcsec\, were recommended for the pn and MOS data, respectively. A circular background region having a radius of 50\arcsec\, was used throughout. The location of the background region was recommended by the \texttt{ebkgreg} tool. The \texttt{rmfgen} and \texttt{arfgen} tools were used to create the response matrix files and ancillary response functions.

\subsubsection{\textit{Chandra}}
\label{sec_chandra}
{\it Chandra} was used to observe \mxb eight times after outburst I and, so far, four times after outburst II. All the observations were carried out in the \texttt{FAINT} mode and the source was positioned on the S3 chip of the Advanced CCD Imaging Spectrometer (ACIS). The data\footnote{Obtained using the {\it Chandra} archive: http://cda.harvard.edu/chaser/} were reduced using \texttt{CIAO} (v4.9). A circular source extraction region having a radius of 2\arcsec\, and an annular background region having an inner and outer radius of 10\arcsec\, and 20\arcsec\, (both centred on the source position) were used throughout, for the light curve and spectra extraction. The data were examined for background flaring (by examining the light curve from the whole field-of-view excluding the region around the source) and only one observation (obsID: 3795) showed such a flare. To correct for this episode of background flaring, data exceeding >4.5 counts s$^{-1}$ were removed from this observation. This reduced the useful exposure time of this observation from $\sim$27.1 ksec to $\sim$25.6 ksec.

The spectra were extracted using the \texttt{specextract} tool. This tool generates the source and background spectrum along with the redistribution matrix file and the aperture corrected auxiliary response file. Two observations (obsID: 5469 and 6337) were very close in time ($\sim$1400 days and $\sim$1417 days after outburst I). We combined the spectra from these two observations to obtain better constraints from our spectral fitting using the \texttt{combine${\_}$spectra} tool.\footnote{This was not done during previously reported analyses of the source \citep{cackett2008cooling}. Using both the combined and non-combined data yield results consistent with one another.} The last two observations after the end of outburst I have not been included in our analysis (see Section \ref{sec_introduction}, for details).

\subsubsection{Spectral Fitting}
All the {\it XMM-Newton} and {\it Chandra} spectra were grouped to have a minimum of 5 counts per bin and the {\it Swift}/XRT Interval 1 spectrum was grouped to have a minimum of 2 counts per bin. The {\it XMM-Newton} spectra were grouped using the \texttt{specgroup} tool and the {\it Chandra} and XRT spectra using the \texttt{grppha}  tool. Due to the low number of counts per bin for the various spectra $\chi^2$ statistics could not be used for the spectral fitting. Therefore, all the spectra were fit collectively using \texttt{XSpec} \citep[v12.9;][]{arnaud1996xspec} in the 0.3--10 keV energy range using W-statistics \citep[background subtracted-Cash statistics;][]{wachter1979parameter}. We fit our spectra using the NS atmosphere model \citep[\texttt{nsatmos;}][]{heinke2006hydrogen}. We assumed a NS mass and radius of 1.6 M$_\odot$ and 12 km\footnote{We have also carried out the spectral fitting and \nscool modelling assuming a NS mass and radius of 1.4 M$_\odot$ and 10 km, as was assumed for \mxb for the post-outburst I cooling studies \citep{cackett2013change}. We find that this does not change the broad physical interpretation of our results.}, since there is evidence that the NS core in \mxb exhibits Direct Urca processes \citep{brown2018rapid} and therefore it may harbour a NS more massive than the canonical 1.4 M$_\odot$ NS. Analysis of the type-I bursts of \mxb for hydrogen-rich and helium-rich material indicated distances of 9$\pm$2 kpc and 12$\pm$3 kpc, respectively \citep{galloway2008thermonuclear}. We assume a distance (D) of 9 kpc.\footnote{We also investigated both, the spectral analysis and the \kteff modelling, assuming D = 12 kpc. The physical interpretation of our results remains the same.} We examined the {\it Gaia} archive and found that the source position coinciding with \mxb was not accompanied by any stellar parallax information. Therefore, no distance constraint could be obtained using the {\it Gaia} data. We assume that the entire NS surface is emitting and set the related normalisation to 1 in the \texttt{nsatmos} model. The equivalent hydrogen column density (\nh) was modelled using \texttt{tbabs}, employing \texttt{VERN} cross-sections and \texttt{WILM} abundances \citep{verner1996atomic,wilms2000absorption}.

We assume that the \nh remained constant throughout and tie it across all the spectra.\footnote{Since we do not include the last two observations after outburst I in our analysis we find (from our study) that this assumption is valid.} The effective NS temperature was left free to vary across all the observations but tied between the pn and MOS detectors for a given {\it XMM-Newton} observation. To account for the normalisation offset between the different observatories we used an additional \texttt{constant} component as has also been used in previous cooling studies \citep[see][for details]{parikh2017different}. The value of this component was determined using Table 5 of \citet[][$C_\text{XRT} = 0.872$, $C_\text{pn} = 0.904$, $C_\text{MOS1} = 0.983$, $C_\text{MOS2} = 1$, and $C_\text{Chandra} = 1$]{plucinsky2017SNR}. No additional non-thermal component was needed to fit the spectra. All errors are stated for the 90 per cent confidence level and all the measured effective temperatures are in terms of the effective surface temperature that would be seen by an observer at infinity\footnote{\kteff = $kT_\mathrm{eff}/(1 + z)$, where $(1 + z)$ is the gravitational redshift factor. For $M_\mathrm{NS} = 1.6 \, M_\odot$ and $R_\mathrm{NS} = $ 12 km, $(1 + z)$ = 1.29.} (\kteff). 

The best-fit \nh was \nh = (3.4$\pm$0.2)$\times 10^{21}$ cm$^{-2}$. The \nh was fixed to this value before calculating the errors on the \kteff to obtain a more constraining result \citep[for justification of this see, e.g.,][]{wijnands2004monitoring,homan2014strongly,parikh2017variable}. The results of the spectral fitting are shown in Table \ref{tab_log} and the \kteff evolution of the cooling crust is shown in Figure \ref{fig_nscool}

\subsection{Modelling the \kteff evolution}
\label{sect_nscool}
\begin{figure}
\centering
\includegraphics[width=9.34cm]{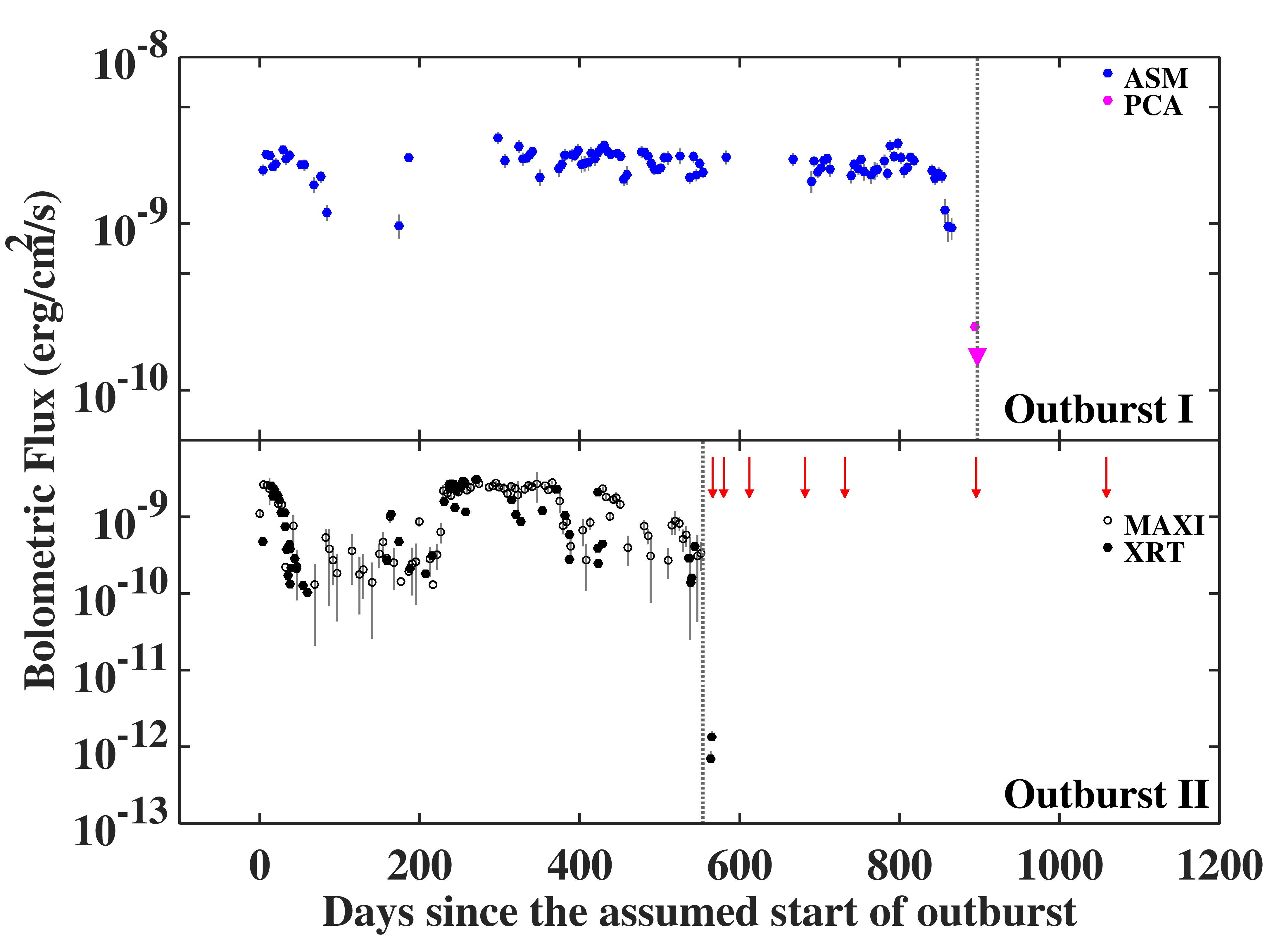}
\caption{The bolometric flux (0.01--100 keV) curves for outbursts I and II are shown in the upper and lower panels, respectively. The zero points correspond to MJD 51265 for outburst I and MJD 57256 for outburst II. The vertical grey dotted lines indicate the time of the end of the respective outbursts (MJD 52162 and MJD 57809.7, respectively). For outburst I, the ASM data is shown in blue and the PCA data near the end of the outburst (including the upper limit indicated by the downward facing triangle) is shown in magenta. For outburst II, the {\it MAXI} and XRT data are shown by open and filled black circles, respectively. The vertical red arrows in the lower panel indicate the times of the observations of the source in quiescence after the end of outburst II (see Section \ref{sect_spectral_analysis} and Table \ref{tab_log}, for details).}
\label{fig_lc}
\end{figure}

\begin{figure}
\centering
\includegraphics[width=9.4cm]{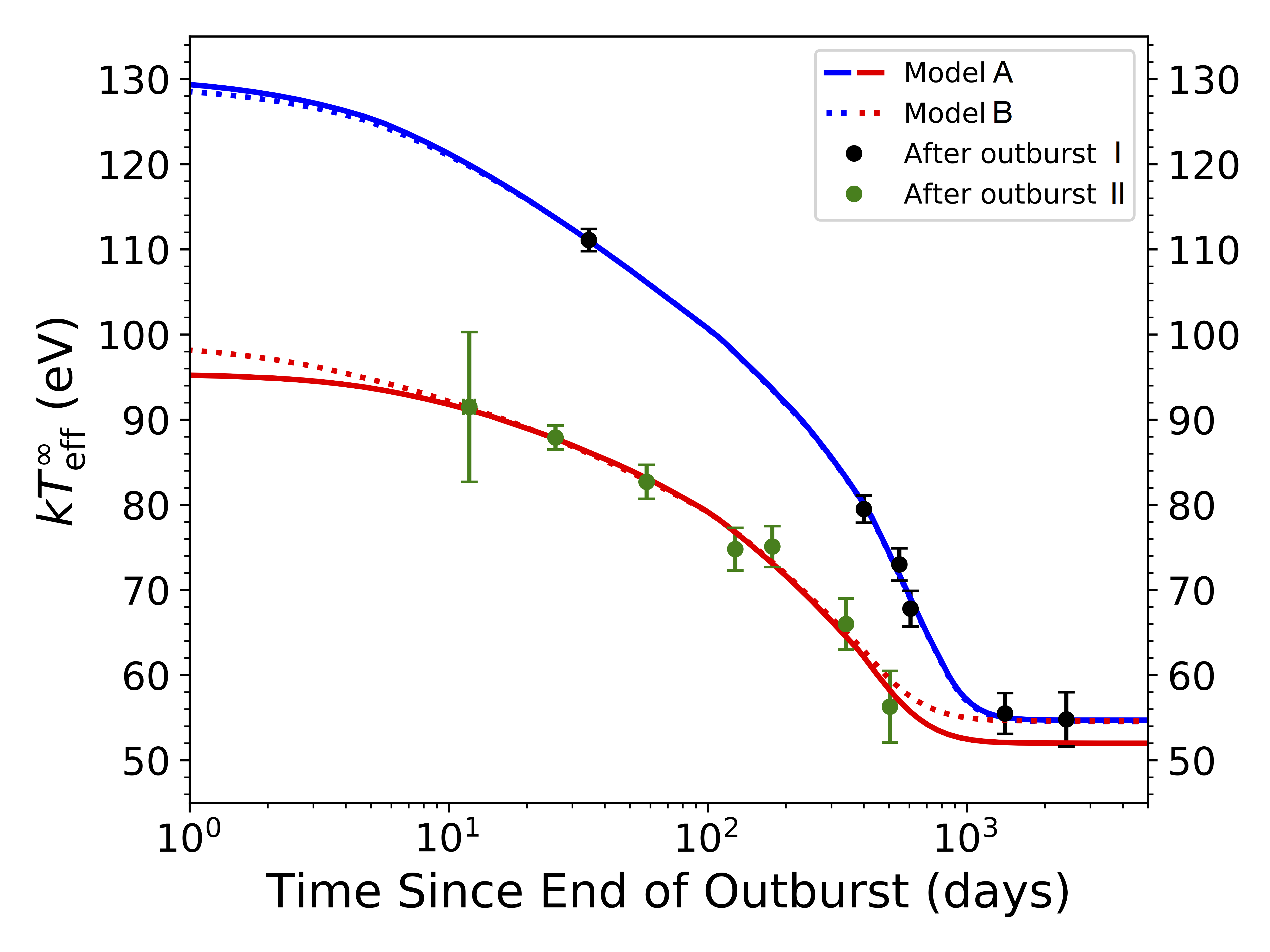}
\caption{The \kteff evolution of \mxb after outbursts I and II is shown by the black and green points, respectively. We have modelled this observed evolution with the crust heating and cooling code \nscool. The modelled cooling curves after outbursts I and II are shown in blue and red, respectively. Model A (shown by the solid lines) indicates the fit when all the parameters were free to vary. Model B (shown by the dotted lines) assumes that y$_\mathrm{light}$ after both the outbursts is the same and, therefore, that the crust returns to the same observed base level. It should be noted that Models A and B have parameters that are consistent with one another within their error bands. This is shown in Figure \ref{fig_nscool_error} and Table \ref{tab_nscool}.}
\label{fig_nscool}
\end{figure}

\begin{figure}
    \centering
    \begin{subfigure}[b]{0.47\textwidth}
        \includegraphics[width=\textwidth]{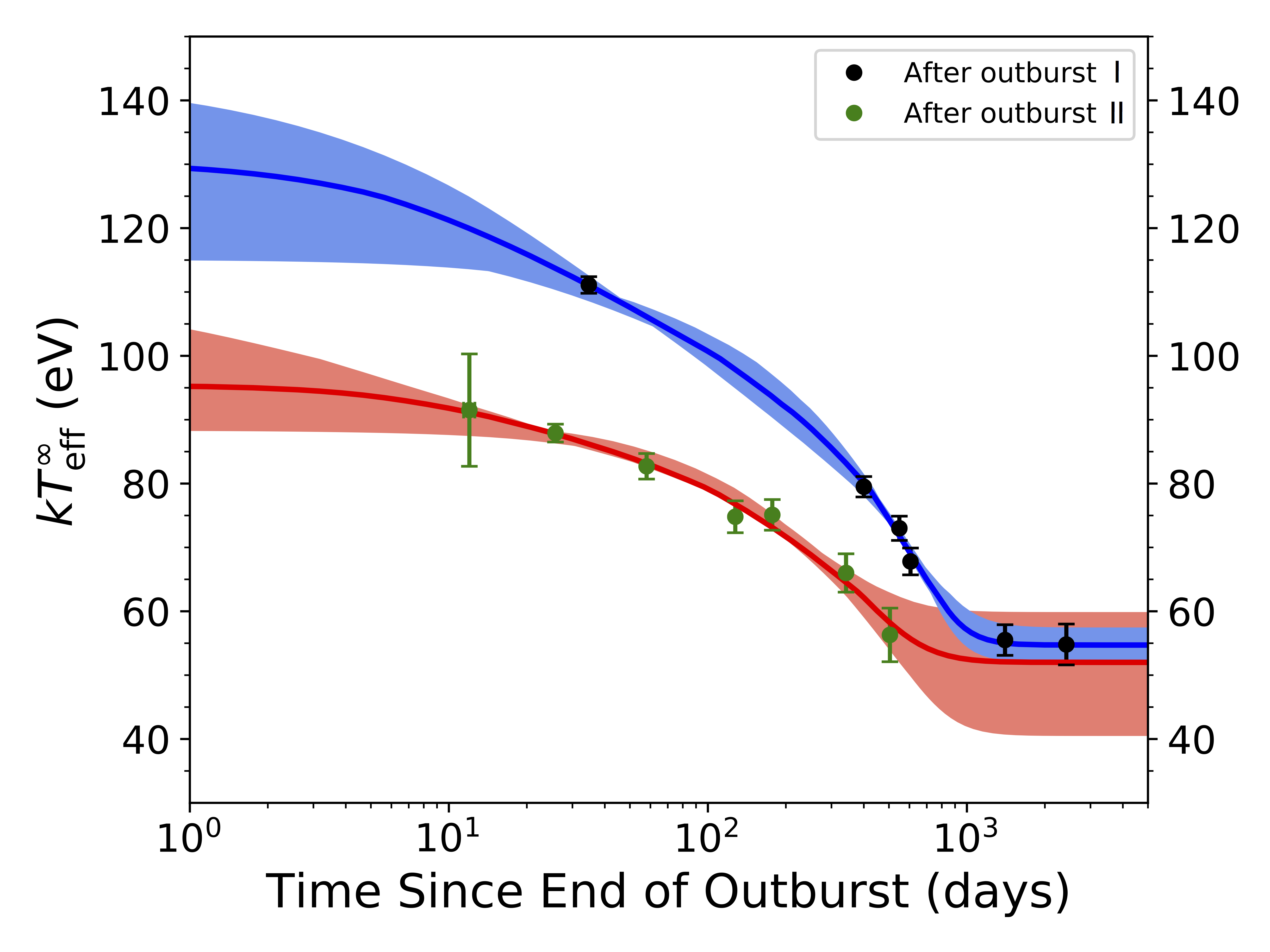}
        \caption{Model A}
        \label{fig_nscool_error_a}
    \end{subfigure}
 
     \begin{subfigure}[b]{0.47\textwidth}
        \includegraphics[width=\textwidth]{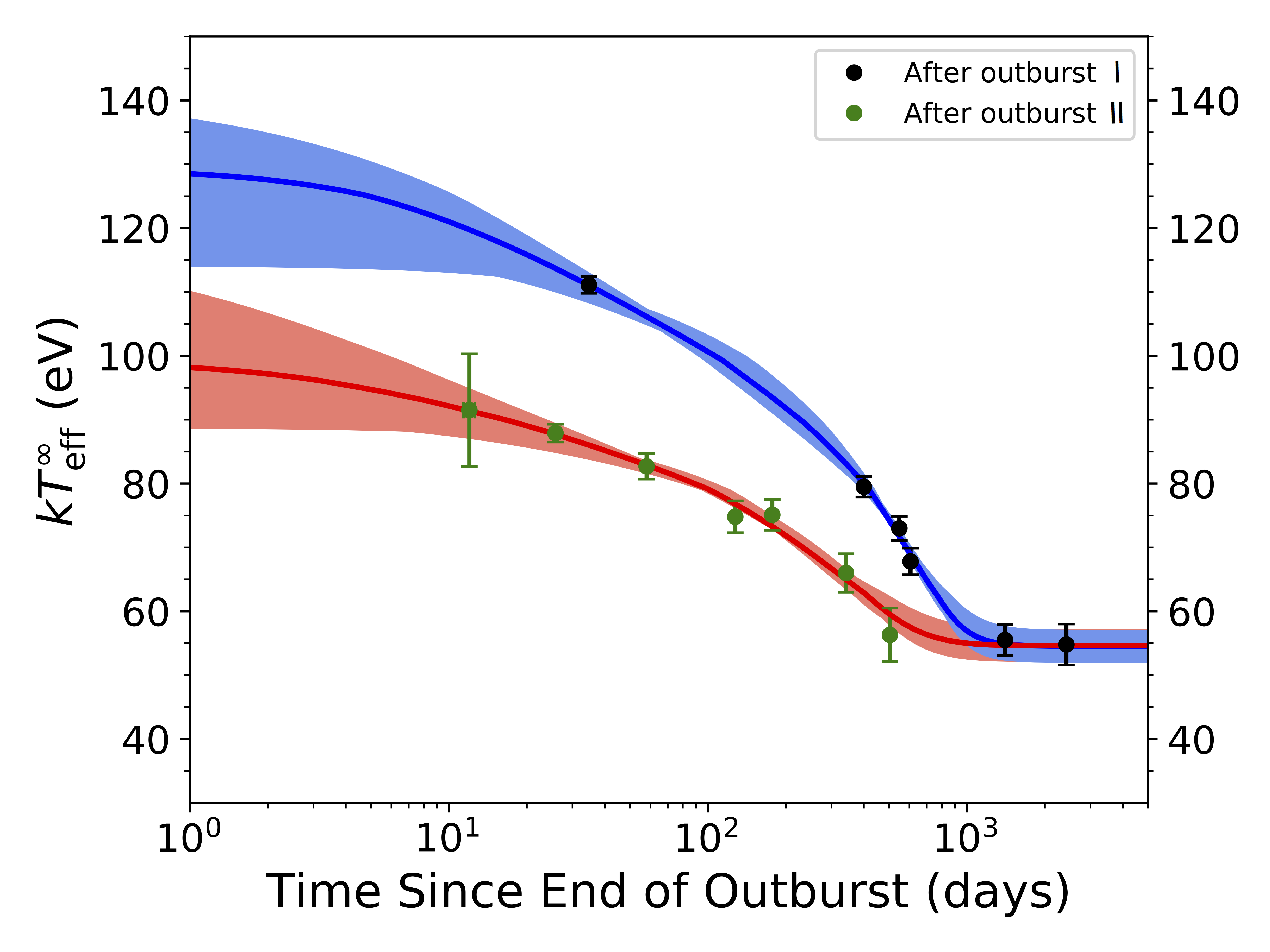}
        \caption{Model B}
        \label{fig_nscool_error_b}
    \end{subfigure}
    \caption{The cooling curves modelling the \kteff evolution of \mxb are shown, along with the error bands on the models. The post-outburst I and post-outburst II \kteff evolution data points are shown in black and green and the modelled cooling curves are shown in blue and red, respectively. Panel a shows Model A for which all parameters were free to vary. Panel b shows Model B for which the y$_\mathrm{light}$ after the end of the two outbursts was assumed to be the same (see also Figure \ref{fig_nscool}). }

\label{fig_nscool_error}
\end{figure}

\begin{table*}
\centering
\caption{The results of the \nscool model fits to the observed \kteff evolution in \mxb after outbursts I and II are shown. The errors are stated for the 90 per cent confidence level. The $\ast$ indicates that the error is not constrained and corresponds to the maximum or minimum allowed value for the given parameter$^\text{a}$.}
\label{tab_nscool}
\small{
\begin{tabular}{lccccccccccl}
\hline
Model & $\tilde{T}_0$ & \Qsh$_\mathrm{,I}$ & \Qsh$_\mathrm{,II}$ & \rhosh$_\mathrm{,I}$ & \rhosh$_\mathrm{,II}$ & log[y$_\mathrm{light,I}$] & log[y$_\mathrm{light,II}$] & \Qimp$_\mathrm{,1}^\text{b}$ & \Qimp$_\mathrm{,2}^\text{b}$ & \Qimp$_\mathrm{,3}^\text{b}$ & $\chi^2$ (d.o.f.$^\text{c}$) \tabularnewline
 & ($\times 10^7$ K) & (MeV & (MeV & ($\times 10^8$ & ($\times 10^8$ & (g cm$^{-2}$) & (g cm$^{-2}$) & & & & \tabularnewline
  & & nucleon$^{-1}$) & nucleon$^{-1}$) & g cm$^{-3}$) & g cm$^{-3}$) &  &  & & & & \tabularnewline 
 \hline 
A & 3.1$^{+1.0}_{-0.5}$ & 1.2$\pm$0.8 & 1.2$^{+2.4}_{-0.7}$ & 4.3$^{+87.5}_{\ast}$ & 10.1$^{+90.1}_{\ast}$ & 8.5$^{\ast}_{-1.7}$ & 7.8$^{+3.0}_{\ast}$ & 2.7$^{+13.7}_{\ast}$ & 2.1$^{+1.9}_{\ast}$ & 1.8$^{+1.6}_{\ast}$ & 2.7 (10)\tabularnewline [0.1cm]
B & 3.2$^{+1.0}_{-0.5}$ & 1.2$\pm$0.7 & 1.0$\pm$0.8 & 5.1$^{+48.5}_{\ast}$ & 3.8$^{+91.7}_{\ast}$ & 8.4$^{\ast}_{\ast}$ & 8.4$^{\ast}_{\ast}$ & 2.3$^{+13.6}_{\ast}$ & 2.1$^{+1.9}_{\ast}$ & 1.6$^{+1.7}_{\ast}$ & 2.9 (9)\tabularnewline [0.1cm]
C & 3.1$^{+1.0}_{-0.5}$ & 1.1$\pm$0.8 & 1.1$\pm$0.8 & 3.5$^{+72.2}_{\ast}$ & 3.5$^{+72.2}_{\ast}$ & 8.6$^{\ast}_{\ast}$ & 7.8$^{+2.6}_{-1.7}$ & 5.9$^{+11.1}_{\ast}$ & 1.7$^{+2.2}_{\ast}$ & 1.9$^{+0.2}_{\ast}$ & 2.6 (8)\tabularnewline [0.1cm]

\hline
\multicolumn{12}{p{17.5cm}}{$^\text{a}$The lowest allowed limit for \rhosh\, is $\rho = 10^{8}$ g cm$^{-3}$ which corresponds to the boundary of the outer crust with the envelope. The highest and lowest allowed y$_\mathrm{light}$ in our model is, respectively, log[y$_\mathrm{light}$] = 12 g cm$^{-2}$ and 5 g cm$^{-2}$. And the lowest allowed \Qimp = 0.}\tabularnewline
\multicolumn{12}{p{17cm}}{$^\text{b}$The \Qimp\, parameters have been indicated as \Qimp$_\mathrm{,\it n}$ where $n$ = the layer of the crust (from the outer crust to the inner crust, n = 1, 2, 3). The outer and inner boundaries of the crust are defined by $\rho = 10^8$ g cm$^{-3}$ and $\rho = 1.5 \times 10^{14}$ g cm$^{-3}$, respectively. The three layers of the crust for which the different \Qimp\, have been modelled are defined by boundaries set at $\rho = 6.2 \times 10^{11}$ g cm$^{-3}$ and $\rho = 8 \times 10^{13}$ g cm$^{-3}$, respectively. Thus, the three layers define the outer crust, the neutron drip layer, and the inner crust where the nuclear pasta is expected to occur.}\tabularnewline
\multicolumn{12}{p{17cm}}{$^\text{c}$Degree of freedom.}\tabularnewline
\end{tabular}
}
\end{table*}

We model the \kteff evolution of \mxb after both outbursts I and II using the crust heating and cooling code \nscool \citep{page2016nscool}. We account for the accretion rate variability during the outbursts in our model by using the observed variability in the bolometric flux \citep[\fbol, 0.01--100 keV;][our code also allows for multiple outburst to be followed in this way; see \citeauthor{parikh2017different} \citeyear{parikh2017different} and \citeauthor{ootes2017} \citeyear{ootes2017} for details]{ootes2016}. To obtain this \fbol, we use the light curves (see Section \ref{sect_lc}, for details) from various instruments and determine appropriate count rate to \fbol\, conversion factors. For outburst I, we use the 2--10 keV {\it RXTE}/ASM light curve and the more sensitive 2--10 keV {\it RXTE}/PCA observations near the end of the outburst. For outburst II, we use the 2--10 keV {\it MAXI}/GSC light curve as well as the 0.5--10 keV {\it Swift}/XRT data.

Recently \citet{iaria2018broadband} reported the \fbol\, of \mxb during high- and low-flux states during outburst II. They also showed that the source likely exhibited the same high-flux state (observed during outburst II) during outburst I as well (MJD 51961 and MJD 57499, respectively; see their Section 2.4). Since the source exhibits the high-flux state during most of both the outbursts we have only used the high-flux \fbol\, in determining our conversion factors for both outbursts I and II. The reported unabsorbed high-flux $F_\mathrm{bol}$ is $2.2 \times 10^{-9}$ erg cm$^{-2}$ s$^{-1}$. This \fbol\, has been corrected for all bursts, eclipses and dipping behaviour and is representative of the persistent emission of the source during the high-flux state.

The count rate to \fbol\, conversion factor for the ASM, {\it MAXI}, and {\it Swift} have been determined using the count rate during the observation performed closest in time to the data from which \citet{iaria2018broadband} obtained the \fbol. We ensure that the count rate corresponding to this observation is representative of the persistent emission from the source (and does not experience any bursts, eclipses, or dipping behaviour). The count rate to \fbol\, conversion factors for the various instruments are: C$_\mathrm{ASM}$ = 1.0$\times$10$^{-9}$ erg cm$^{-2}$ counts$^{-1}$, C$_\mathit{MAXI}$ = 2.6$\times$10$^{-8}$ erg cm$^{-2}$ counts$^{-1}$ and C$_\mathit{Swift}$ = 1.4$\times$10$^{-10}$ erg cm$^{-2}$ counts$^{-1}$. A similar factor could not be determined for the PCA data near the end of outburst I since these data were not coincident with the time of the \fbol\, reported during this outburst. Instead, we used a correction factor of 2 \citep{zand2007six} to convert the 2--10 keV flux to the \fbol. These \fbol\, curves are shown in Figure \ref{fig_lc}. The  upper panel shows outburst I with the 4-day binned and error filtered ASM data shown in blue and the PCA data (including the upper limit shown by a downward pointing triangle) in magenta. The lower panel shows the bolometric flux curve of outburst II with the {\it MAXI} data shown using the open black circles and the {\it Swift}/XRT data by the filled black circles. For outburst II, in case when both {\it MAXI} and XRT data were available for the same day the XRT data were preferred. The vertical dotted grey line in both the panels indicates the end of the outburst. The vertical red arrows in the lower panel show the times of the quiescent observations \citep[see Section \ref{sect_spectral_analysis} and Table \ref{tab_log}; see][for the times of the quiescent observations after the end of outburst I]{cackett2006cooling,cackett2010continued}.

These calculated \fbol\, curves were then used to determine the daily average accretion rate using 
\begin{equation}
\centering
\label{eqn_m_dot}
\dot{M} = \frac{F_\mathrm{bol} 4 \pi D^2}{\eta\, c^2}
\end{equation}
where $\eta$ (= 0.2) indicates the efficiency factor and $c$ is the speed of light. These \fbol\, curves were used to calculate the outburst fluence. The fluence for outbursts I and II were found to be $\sim$0.17 erg cm$^{-2}$ and $\sim$5.18$\times$10$^{-2}$ erg cm$^{-2}$, respectively.

For consistency in our \nscool modelling, we used the same values of the NS mass and radius as assumed for the spectral analysis (see Section \ref{sect_spectral_analysis}). Data after both outbursts were fitted collectively to obtain the best model constraints. Both, the deep crustal heating as well as shallow heating were modelled. The contribution of deep crustal heating was fixed at 1.93 MeV nucleon$^{-1}$ \citep[][]{haensel2008models}. The strength (\Qsh) and the depth (\rhosh) of the shallow heating were model fit parameters. The additional model fit parameters were the column depth of light elements in the envelope\footnote{The envelope constitutes the outer NS where $\rho < 10^8$ g cm$^{-3}$ and its composition determines how the temperature at the bottom of the envelope is translated to a surface temperature which is then measured by the observer \citep[see also][]{ootes2017}.} (y$_\mathrm{light}$), the initial red-shifted core temperature ($\tilde{T}_0$) of the NS, and the impurity factor of the crust (\Qimp) which was modelled as three layers.\footnote{These layers correspond to the outer crust, the neutron drip layer, and the inner crust where the nuclear pasta is expected to occur \citep[see][for details, as well as footnote b of Table \ref{tab_nscool}]{ootes2017}.} Our model assumes that the \Qimp\, in the different crustal layers remains the same between the two outbursts. The best fit was found using the $\chi^2$ minimisation technique. 

Initially, we allowed the shallow heating parameters (\Qsh\, and \rhosh) and y$_\mathrm{light}$ to vary between the two outbursts, as it was found in studying multiple cooling curves of other sources that these parameters could indeed be different between outbursts \citep{parikh2017different,ootes2017}. This is shown as Model A in Figure \ref{fig_nscool} (by the solid curve; blue and red are used to indicate the cooling curves after outbursts I and II, respectively) and Table \ref{tab_nscool}. The fit indicates that the \Qimp\, in all the layers is low, with the best-fit indicating that \Qimp$\lesssim$3 for all the three layers (with the lowest bound extending to 0). The fit values of the shallow heating parameters (active during the accretion outburst) and y$_\mathrm{light}$ (at the end of the outburst) during the collective modelling of the two outbursts are consistent with one another. The y$_\mathrm{light}$ is used to translate the boundary temperature at $\rho = 10^8$ g cm$^{-3}$ into the temperature seen by an observer \citep[see Figure 1 of][]{ootes2017}. Since the best-fit y$_\mathrm{light}$ for Model A is different after outbursts I and II the source reaches a different base level (as can be seen from Figure \ref{fig_nscool} after $\gtrsim$1000 days). However, the y$_\mathrm{light}$ values are consistent within the error bars and may still be the same after the two outbursts (as can be seen in Figure \ref{fig_nscool_error_a} which shows the cooling curves of Model A along with its error bands). We can investigate this further using our upcoming {\it Chandra} observations. We have also modelled the source reaching the same base level (once the crust returns to equilibrium with the core) after both the outbursts as Model B (shown by the dotted curve in Figure \ref{fig_nscool} and with its corresponding error band in Figure \ref{fig_nscool_error_b}) by tying the y$_\mathrm{light}$ between the two outbursts. All fit parameter values between Models A and B are consistent (see Table \ref{tab_nscool}). 

We have also modelled the cooling evolution of the source assuming that the shallow heating parameters between the two outbursts are the same (as Model C, see Table \ref{tab_nscool}). This gives, as expected, a similar result as seen for Model A, i.e., y$_\mathrm{light}$ between the two outbursts is consistent and the \Qimp\, is low. We do not show Model C in Figure \ref{fig_nscool} because it almost entirely overlaps with Model A. 

Using our \nscool model, we have also examined the evolution of the temperature profile in the NS crust during and after the end of both accretion outbursts in MXB 1659$-$29. This has been done for Model B, presented as Video 1 (see online material). Outburst I is indicated using blue and outburst II using red. The left panel shows the mass accretion rate variability during both the outbursts (shown as the effective temperature as a function of time). The upper left and lower left panel show outbursts I and II, respectively and the zero point (shown by the vertical dotted line) indicates the time of transition to quiescence. The right panel shows the temperature profile in the neutron star and dashed vertical line is indicative of the crust-core boundary. The temperatures in the right panel are the local, i.e., non-redshifted, temperatures. The video has been presented such that both outbursts (although they have different lengths) transition to quiescence at the same time. This means that the accretion in outburst I starts before that in outburst II as outburst I is longer. 

The temperature profile in the crust of a neutron star is set by the core temperature when the crust is in thermal equilibrium with the core. In our system, this crust-core equilibrium condition exists before and after the two accretion outbursts. The pre-outburst II profile is the profile in the crust {\it after} the end of outburst I (once the crust-core equilibrium has been re-established). In our model we have assumed that the core is not superfluid. 

As the source begins to accrete the crust begins to heat up (seen as two bumps in the temperature profile) due to the two heat sources --- the deep and shallow crustal heating. Very quickly this heat spreads through the crust yielding a much smoother temperature profile. The profile in the outer crust is very variable as the shallow heat source responds nearly instantaneously to the outburst accretion rate variability. 

The bump around the depth where deep crustal heating occurs is lower for outburst II than for outburst I as outburst II was shorter (with a smaller total accreted mass). It can be seen that the crust begins to cool as soon as the accretion stops. As seen from our observations, outburst I takes longer to achieve equilibrium with the core than outburst II. 
   
\section{Discussion}
\label{sec_disc}
\mxb is a LMXB that hosts a NS. We study the crust cooling of the NS crust in \mxb after two accretion outbursts using data obtained from the {\it Swift}, {\it XMM-Newton}, and {\it Chandra} observatories. The two outbursts studied here --- outburst I (1999--2001) and outburst II (2015--2017) have a similar peak flux but had different lengths, lasting $\sim$2.5 years and $\sim$1.7 years, respectively. Experience from the previous outburst and theoretical expectations showed the importance of observations during the early cooling phase soon after the end of the outburst. For MXB 1659$-$29, we have obtained a much improved coverage for the first $\sim$200 days after the end of outburst II as compared to outburst I.

We reduced and modelled all the quiescent crust cooling data collectively for consistency. We found that during outburst I the crust was heated up to higher temperature than during outburst II, but after both outbursts the NS crust exhibited cooling. The post-outburst cooling, as inferred from our spectral fitting results, indicated a \kteff drop from $\sim$111 eV to $\sim$55 eV from $\sim$36 days to $\sim$2422 days after the end of outburst I and from $\sim$92 eV to $\sim$56 eV from $\sim$12 days to $\sim$505 days after the end of outburst II. The \kteff extracted from the most recently performed observation after outburst II is consistent with the \kteff that was assumed to represent the crust-core equilibrium after outburst I (see Table \ref{tab_log} and Figure \ref{fig_nscool}). This suggests that the crust may be close to returning to thermal equilibrium with the core if the assumed base level after both outbursts is the same (see Model B). We will obtain at least two more {\it Chandra} observations of the source in the future (currently planned for 2019). This will provide us with information about whether the crust will cool further or not since it has re-established thermal equilibrium with the core. 

Future observations in quiescence may also help break the ambiguity of the last observations after the end of outburst I \citep[taken $\sim$3 days apart, see Section \ref{sec_introduction} and][]{cackett2013change}. Examining high quality spectra taken at a similar time $\sim$11 years after the end of outburst II (if the source does not start a new accretion outburst before that) will allow us to infer if the base level does drop or if indeed the \nh increases due to build up of material in the disk.

We have collectively modelled the cooling trend observed after outbursts I and II using our theoretical crust heating and cooling code \nscool \citep{page2016nscool,ootes2016,ootes2017}. We have assumed that the impurity parameter \Qimp\, in the crust does not vary between the outbursts. As is seen for several sources \citep{shternin2007neutron,brown2009mapping,page2013forecasting,ootes2016}, our models indicate a low \Qimp\, (best-fit shows \Qimp$\lesssim$6) in the crust, meaning a high thermal conductivity. Initially, we allowed both the shallow heating parameters (\Qsh\, and \rhosh) and the envelope composition (y$_\mathrm{light}$) to vary between both the outbursts. Our models suggest that all our \nscool fit parameters are consistent between the two outbursts and none of the parameters needed to be adjusted/varied between the two cooling curves. This makes \mxb a predictable cooling source whose quiescent cooling evolution after a new outburst can be calculated using information obtained from the cooling curve after a previous outburst. \citet[][see their Figure 1, right]{wijngaarden2017window} reported the early post-outburst II cooling results of this source (from Interval 1 determined using the {\it Swift}/XRT data and the first {\it XMM-Newton} observation after the end of outburst II). We have obtained five additional observations of \mxb since then. It is very interesting to note that the \kteff determined from the five subsequent cooling observation are nicely consistent with the predicted cooling curve we showed in \citet{wijngaarden2017window}. This prediction still holds true when the models are updated for our re-evaluated assumptions (see below) as the updated assumptions result in a revised contribution from the shallow heating which dominates the crust cooling behaviour observed so far after outburst II with very little influence from the deep crustal heating.

Our results, showing that \mxb needs similar shallow heating during both its accretion outbursts, are not consistent with the results reported by \citet{wijngaarden2017window}. This inconsistency is a result of the different \fbol\, assumed by  \citet{wijngaarden2017window} for outburst I. The \fbol\, is used to determine the daily average accretion rate (see Equation \ref{eqn_m_dot}) which in turn is used to determine the \Qsh\, \citep[which is assumed to be proportional to the accretion rate, see Equation 1 of][]{ootes2017}. The  \fbol\, assumed by \citet[][which was determined using WebPIMMS\footnote{https://heasarc.gsfc.nasa.gov/cgi-bin/Tools/w3pimms/w3pimms.pl}]{wijngaarden2017window} for outburst I was a factor of $\sim$2 lower than what we derived using the assumptions indicated in Section \ref{sect_nscool}. Thus, this explains why \citet{wijngaarden2017window} found that outburst I needed a \Qsh\, a factor of $\sim$2 higher than what we find for our model. Our assumption is more robust since it uses actual spectral fit values reported by \citet{iaria2018broadband} to determine the \fbol.

Since the \Qsh\, parameters for \mxb are consistent between the two outbursts and both outbursts have very similar peak fluxes, we have investigated the possibility that the \Qsh\, may be related to the peak flux. To improve the statistics of our study, we have only used the results of MAXI J0556$-$332 and Aql X$-$1 \citep{parikh2017different,ootes2017} because these are the only other two sources for which multiple cooling curves have been collectively modelled. Studying these data we do not find any conclusive evidence that the \Qsh\, may be related to the peak flux of the outburst. 

Using our \nscool model we calculated the fluence of the two outbursts (see Section \ref{sect_nscool}). We find that the fluence of outburst I was a factor of $\sim$3.3 higher than that of outburst II. Even though the two outbursts of \mxb exhibited a different fluence our modelling indicates that they need a similar \Qsh. This is different from the shallow heating requirements of MAXI J0556$-$332 which indicated that different amounts of \Qsh\, were required during its three accretion outbursts to explain their post-outburst cooling evolution and that the magnitude of the \Qsh\, required seemed to be proportional to the outburst fluence \citep{parikh2017different}. Our results are also different from those published by \citet{wijngaarden2017window}. They reported that the \Qsh\, was also proportional to the outburst fluence for MXB 1659$-$29. However, this is no longer true since (as we show earlier in this section) we have robustly recalculated the fluence from \mxb and remodelled the data resulting in a different contribution from the \Qsh. Both the fluence ane the \Qsh\, were found it to be different from those of \citet{wijngaarden2017window}. This results in the \Qsh\, no longer being proportional to the fluence for MXB 1659$-$29.

In our model, the total contribution from the shallow heating is assumed to be proportional to the outburst \mdot variability, with each accreted nucleon contribution \Qsh\, MeV of heat. For MXB 1659$-$29, we have found that the magnitude of heating needed per accreted nucleon (\Qsh\,) was consistent during two different accretion outbursts. This shows that our assumption, of the dependence of total shallow heating on the \mdot variability is robust. However, this is only true for MXB 1659$-$29. MAXI J0556$-$332, another crust cooling source, needed different \Qsh\, during its three outbursts to explain the cooling evolution observed in quiescence.

The origin and nature of the unknown shallow heat source remains a puzzle. However, studying more sources, including different outbursts of the same source, will increase the known constraints on the shallow heat source. Although a slow process, this currently seems one of the only two proven ways forward \citep[the other being studies using type-I bursts; e.g.,][]{cumming2006long,linares2012millihertz,zand2012superexpansion,meisel2018nuclear} in allowing us to infer the physical origin of the shallow heat source. It is important to continue both these complementary studies to ensure that the \Qsh\, requirements are consistent.

\begin{acknowledgements}
AP, RW, LO, and ARE are supported by a NWO Top Grant, Module 1, awarded to RW. ND is supported by an NWO Vidi grant. DP is partially supported by the Consejo Nacional de Ciencia y Tecnolog{\'\i}a with a CB-2014-1 grant $\#$240512. This work benefitted from support by the National Science Foundation under Grant No. PHY-1430152 (JINA Center for the Evolution of the Elements).
\end{acknowledgements}

\bibliographystyle{aa}

\end{document}